\documentstyle[11pt,newpasp,twoside,epsf]{article}
\def\edcomment#1{\iffalse\marginpar{\raggedright\sl#1\/}\else\relax\fi}
\reversemarginpar

\begin{document}
\title{The formation of asymmetries in Multiple Shell Planetary Nebulae due
      to interaction with the ISM}
\author{Eva Villaver, Arturo Manchado}
\affil{Instituto de Astrof\'{\i}sica de Canarias, C/V\'{\i}a L\'actea, 38200
      La Laguna, Tenerife, Spain}
\author{Guillermo Garc\'{\i}a-Segura}
\affil{IA-UNAM, Apdo Postal 877, Ensenada, 22800 Baja California, M\'exico}

\begin{abstract}

Multiple shell planetary nebulae (MSPNe) are most likely a consequence of the
modulation of the mass-loss rate during the thermal pulses on the AGB.
By using numerical simulations of their formation, the models' predictions and
the history of the winds can be investigated. As their halos have low densities
and expansion velocities,  MSPNe are expected to be affected by interaction
with the ISM .  This would then give us evidence of the local ISM conditions.

In order to study the formation of MSPNe,  we have performed numerical
simulations following the evolution of the stellar winds for a 1M$_\odot$ star
from the AGB to the post-AGB stages. Without invoking any asymmetry for the
stellar wind and taking
into account the effects of a moving central star, an asymmetric halo is
formed as a consequence of the interaction with the ISM.

We found that the asymmetries caused by the interaction take place
from the beginning of the evolution and have an
enormous influence on the formation of the halo.
\end{abstract}

\section{Introduction}
The existence of faint shells
surrounding PNe was pointed out by Duncan (1937), and
Chu, Jacoby \& Arendt (1987) studied and classified a large sample of
MSPNe. Since then, the use of more
sensitive CCD detectors and the HST telescope have revealed that these
structures are more common than was originally believed. MSPNe appeared in 24\%
of a complete sample of  spherical
and elliptical PNe in the
northern hemisphere (Manchado 1996); 40\% of these show asymmetries
in the halo that could
be related to the interaction with the ISM.
Many studies have been carried out to try
to establish a connection between the central star evolution and the observed
shells (Stanghellini \& Pascuali 1995, Trimble \&
Sackmann 1978, Frank, van der Venn \& Balick 1994). We have approached
the problem from a different angle:  to find out whether the predictions
of discrete enhanced mass-loss rates during the AGB for low mass stars given
by Vassiliadis \& Wood (1993) are able to reproduce the observed MSPNe.

\subsection{The MSPNe}
We worked out
numerical simulations using ZEUS-3D as a hydro-code and the stellar
evolutionary models as the inner
boundary conditions. We set up the time-dependent wind parameters using the
models of
Vassiliadis \& Wood (1993) during the AGB and Vassiliadis \& Wood (1994) for
the post-AGB stages. During the transition time we adopted a linear
interpolation  between the wind values at the end of the AGB and wind values
at the beginning of the post-AGB. In fig.1. we show the observed H$\alpha$
emission brightness profile of the PN NGC 6826 and the computed ones at
different evolutionary times.

\begin{figure}[ht]
\plottwo{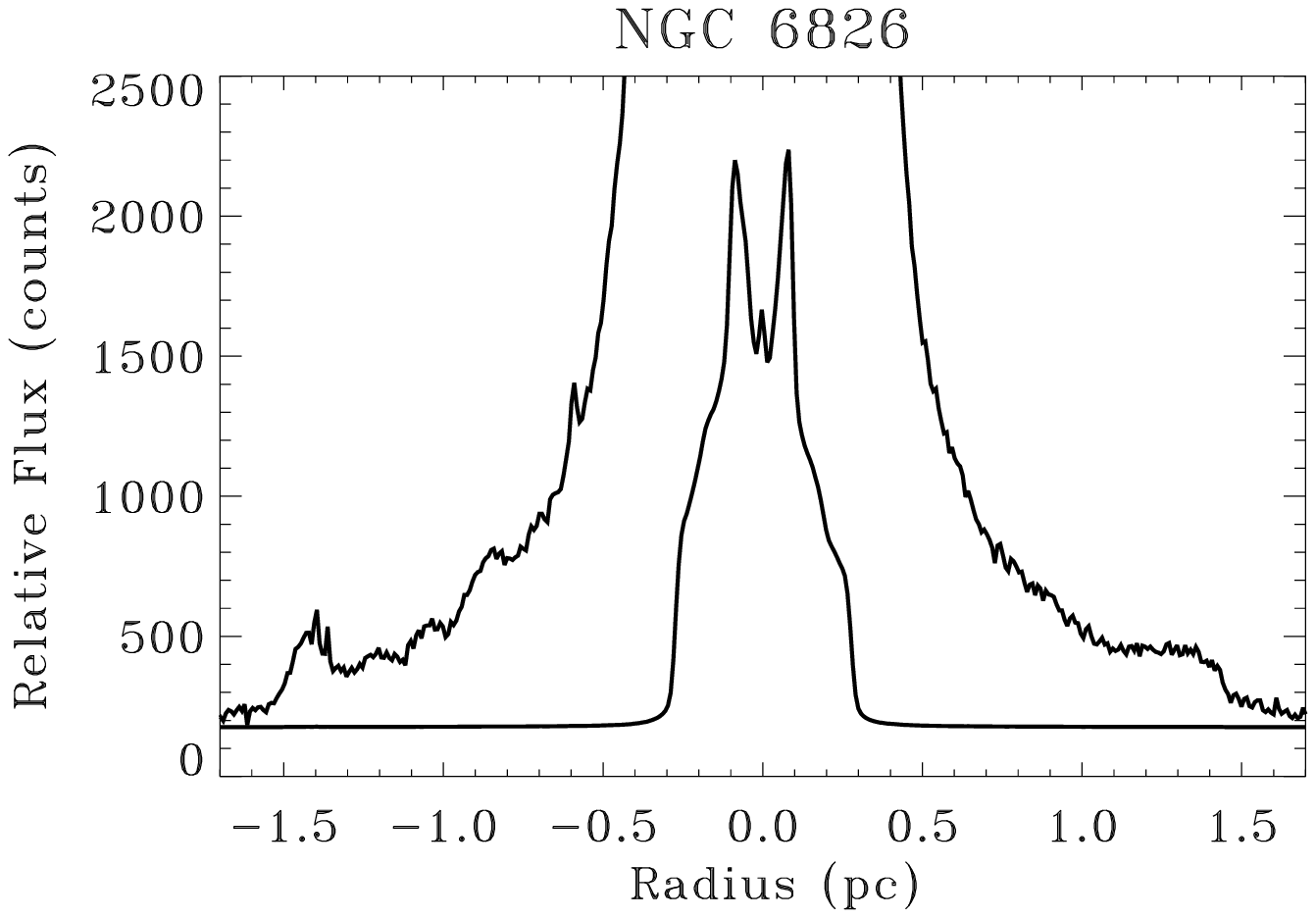}{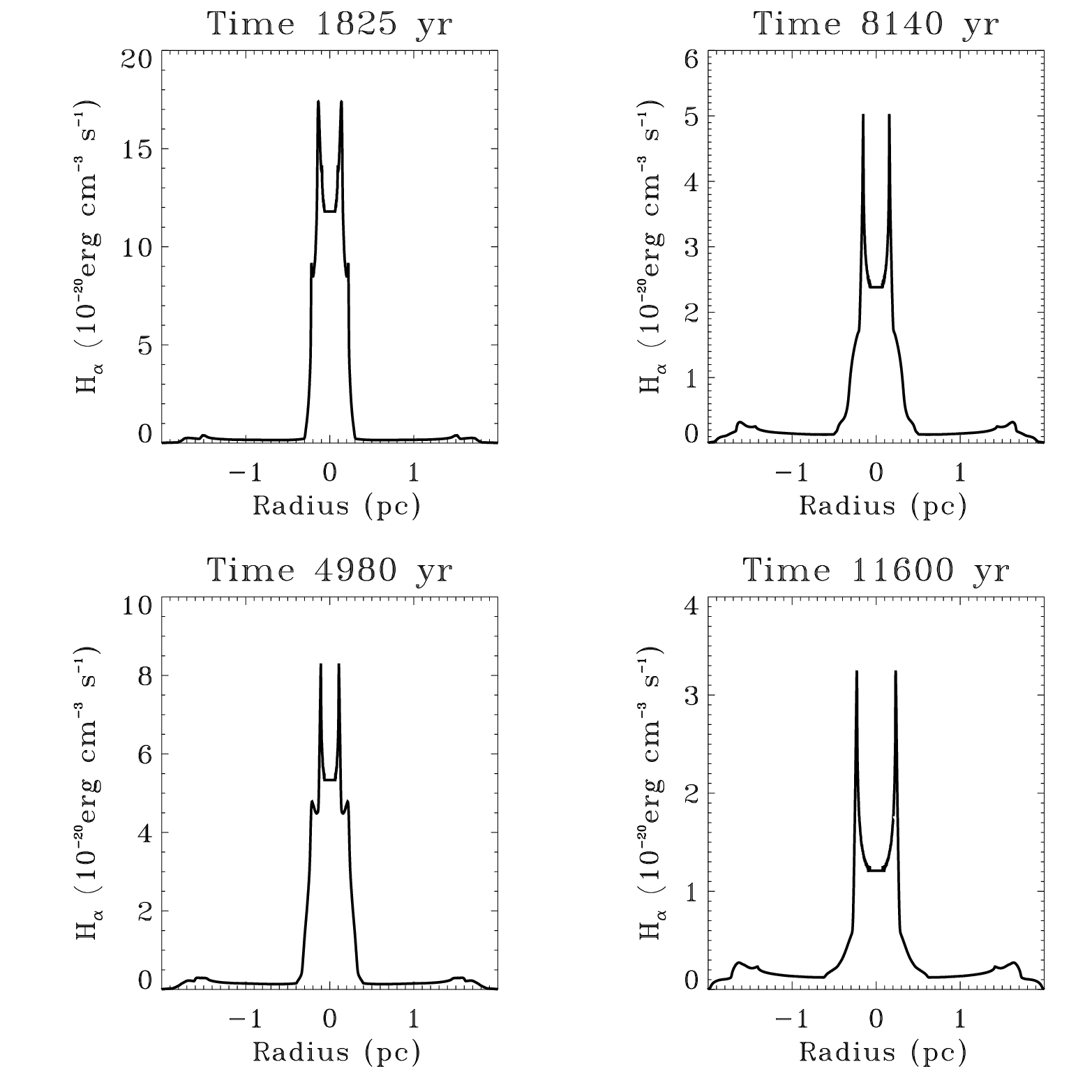}
\caption{On the left H$\alpha$ observed brightness profiles across the
central part of NGC 6826, the central part is also shown but scaled
accordingly. On the right computed H$\alpha$ brightness profiles at different
evolutionary times. Time zero is defined as the time where photoionization
begins.}
\label{fone}
\end{figure}
The halo brightness profiles are characterized by a continuous
decline in the emission and a relative maximum at the edge caused by an
abrupt enhancement of the density on the leading surface of the shell.
The linear size of NGC 6826 has been computed by adopting the spectroscopic
distance of 2.2 Kpc given by Mendez, Herrero \& Manchado (1990).
A direct comparison of the observed and computed profiles shows that
our simulations are able to reproduce the overall shape and size of the nebula.

\subsection{The interaction process}
\begin{figure}
\plotone{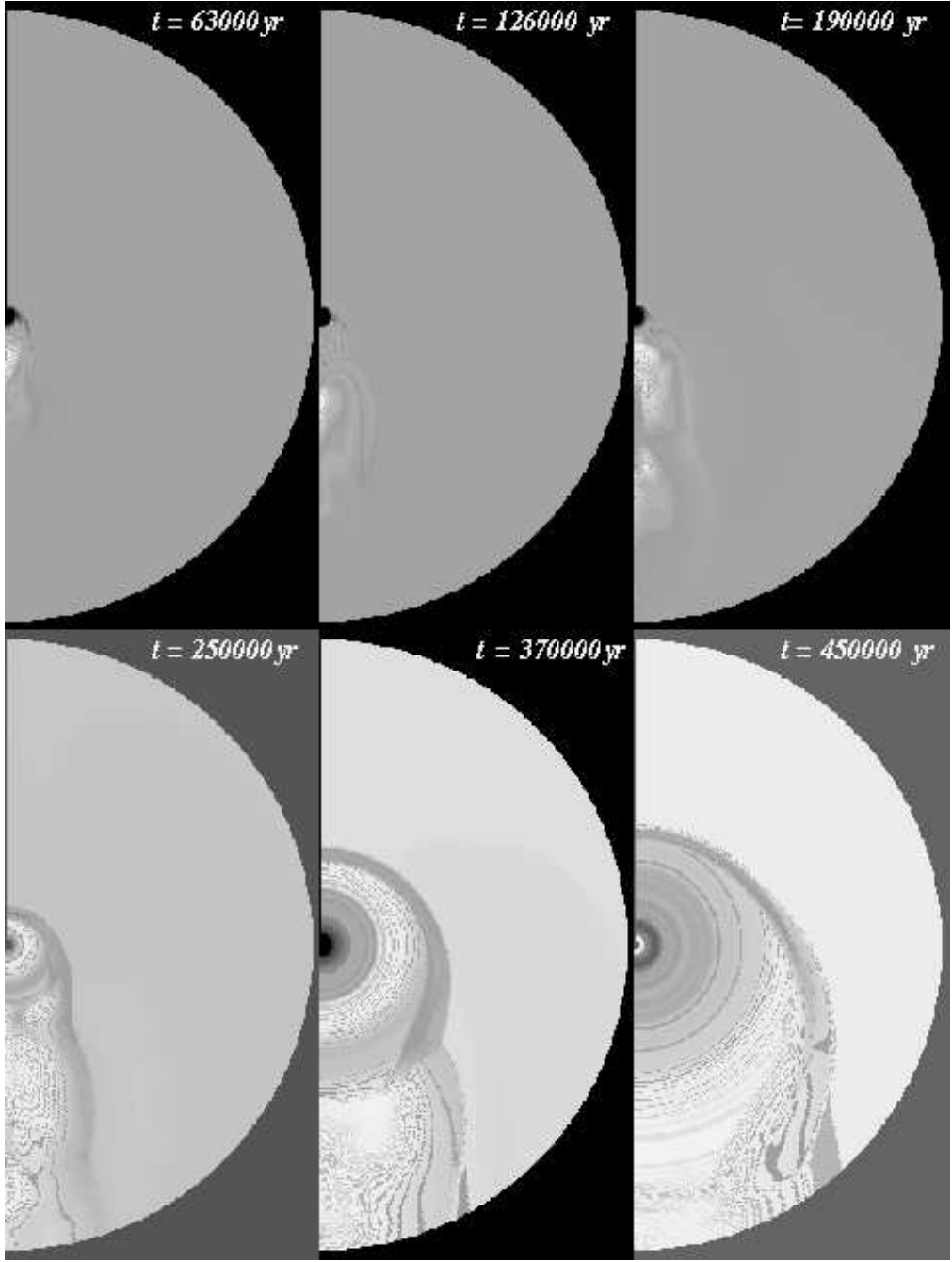}
\caption{The figure shows the log of density at different evolutionary times
of the shell generated by a
star moving with a velocity of 20 $km~s^{-1}$ through a interstellar medium of
density n$_o$=0.1 cm$^{-3}$.
 }
\label{fone}
\end{figure}
The time spent by the star as it evolves during the AGB has been
neglected in previous works related to the study of the interaction process
(Borkowski, Sarazin \& Soker 1990, Soker, Borkowski \& Sarazin 1991).
To address this question we set up the time dependent wind parameters within
a small spherical input
region centered on the symmetry axis, where reflecting boundary conditions
have been used. An outflow boundary condition has been set at the outer
radial direction.

Interaction with the ISM (Figure 2) has been studied assuming that the star
is moving
supersonically with 20 $km~s^{-1}$  across an homogeneous external medium
with density 0.1 $cm^{-3}$.  Isothermal sound speed of the unperturbed ISM is
$c=3~km~s^{-1}$, which give us a Mach number of 7.
The movement takes place perpendicular to the line of sight. The computation
was performed in a 2D spherical grid with the angular coordinate ranging from
0 to 180 degrees.
The interaction between the wind expelled by the star and the ISM takes place
from the beginning of the evolution. A bow-shock configuration is quickly
established (our Mach number, 7,  is  enough to form a strong shock). The
compression across a strong isothermal shock depends on the upstream Mach
number. To keep the gas isothermal, the internal energy has to be
radiated away. This internal energy would otherwise have limited the
compression.
Mass-loss rate associated with the last thermal pulse interacts directly with
the local unperturbed ISM,  giving rise to a less dense shell than the one
formed by a steady star.

\section{Conclusions}
As a consequence of the evolution of a 1M$_\odot$ from the models of
Vassiladis \& Wood (1993,1994) a Multiple Shell Planetary Nebulae is formed.
Since our assumptions of the velocities and the ISM conditions are very
conservative, we can conclude that we will see a spherical halo only if the
star is at rest in relation to the ISM or if it is moving at low angles in
relation to the line of sight.

\section*{Acknowledgment}
We thank M. L. Norman and the laboratory for Computational Astrophysics for
the use of ZEUS-3D.  The work of EV and AM is supported by a grant from the
Spanish DGES PB97-1435-C02-01.

\end{document}